\documentclass[12pt]{article}
\usepackage[latin1]{inputenc}
\usepackage[]{epsfig}
\usepackage{amsfonts}
\usepackage{amsmath}
\usepackage{color}
\usepackage{float}

\newcommand{\be}{\begin{equation}}\newcommand{\ee}{\end{equation}}
\newcommand{\ba}{\begin{eqnarray}}
\newcommand{\ea}{\end{eqnarray}}

\newcommand{\bea}{\begin{eqnarray}} \newcommand{\eea}{\end{eqnarray}}

\def\vr{{\check e}_r}
\def\vteta{{\check e}_\theta}
\def\vfi{{\check e}_\varphi}

\begin{document}

\title{Black Holes in Magnetic Monopoles with a Dark Halo}
\author{A.~Lugo, J.M.~P\'erez Ipi\~na, F.A. Schaposnik
\\
~
\\
{\normalsize  \it Departamento de F\'\i sica,
 Universidad
Nacional de La Plata}\\ {\normalsize\it Instituto de F\'\i sica La Plata-CONICET}\\
{\normalsize\it C.C. 67, 1900 La Plata,
Argentina}
 }
\date{\today}
\maketitle

\begin{abstract}
We study a spontaneously broken Einstein-Yang-Mills-Higgs model coupled via a Higgs portal to an uncharged scalar $\chi$. We present a phase diagram of self-gravitating solutions showing that, depending on the choice of parameters of the $\chi$ scalar potential and the Higgs portal coupling constant $  \gamma$, one can identify different regions:  If $\gamma$ is sufficiently small   a $\chi$ halo is created around the monopole core which in turn surrounds a black-hole. For larger values of $\gamma$ no halo exists and the solution is just a black hole-monopole one. When the horizon radius grows and becomes larger than the monopole radius  solely a black hole solution exists.  Because of the presence of the $\chi$ scalar  a bound for the Higgs potential  coupling constant exists and  when  it is  not satisfied, the vacuum is unstable and no non-trivial solution exists. We briefly comment on possible connections of our results with those found in recent dark matter axion models.\end{abstract}
\section{Introduction}
In this work we analyze a model in which an uncharged scalar $\chi$ is coupled to a Yang-Mills-Higgs system interacting with gravity, looking for solutions in which there is a competition between the extra $\chi$ field and the Higgs   vacuum  expectation value.

Adding   scalars to theories in which gauge symmetry breaking is triggered  by the Higgs mechanism has been  investigated in different contexts.
Indeed, when the  additional uncharged scalar  has a nontrivial v.e.v. in the core of vortices  and monopole solutions,   striking effects take place  in connection with superconducting superstrings  \cite{Witten} and also with the existence of non-Abelian moduli localized on their world sheets \cite{Shifman} -\cite{nosotros}. Also
in the study
of the so called  hidden sector coupled to the Standard Model received much attention in connection with the dark matter problem. In this respect  a possible way to couple the visible and hidden sectors is via the Higgs portal in which  the Higgs field couples with (hidden) $\chi$ scalars  \cite{HiggsPortal} (for a complete list of references see ref.\,\cite{Paola}).

The presence of an additional order parameter associated to the uncharged  scalar has been also discussed in condensed matter models of High-Tc superconductivity. In that case  the original order parameter  in the classical static Landau-Ginzburg free energy density functional is suppressed in the core of vortices while the  competing order associated to an additional  scalar $\chi$ can create  a halo about the vortex core. Depending on the parameter coupling ranges this  leads  to remarkable zero temperature phase diagrams that can be relevant to the description of topological superconductors \cite{Z}-\cite{F2}.

't Hooft-Polyakov  monopole solutions of the $SU(2)$ (or $O(3)$) Yang-Mills-Higgs  model \cite{tH}-\cite{P} also coupled to an additional competing scalar exhibit a similar behavior.  Indeed, adding to this model an $O(3)$ triplet,  uncharged under the gauge group, new collective isospin collective coordinates leads to an   orientational spin model which, after quantization   can be interpreted as a dyon with isospin \cite{Tallarita}.

Concerning monopoles  interacting with gravity, hairy black-hole solutions were  constructed in refs.\,\cite{GV}-\cite{Bizon}. In order to be stable, black holes should be microscopically small since otherwise they will become hairless (see ref.\cite{Volkov} for a complete list of references). In particular, the case of AdS space  holographic phase transitions were discussed in ref.\,\cite{LMS1}.

In this paper we shall consider an $SU(2)$ gauge theory with spontaneous symmetry breaking coupled via  a Higgs portal to a real scalar $\chi$ in a curved space-time to see whether the presence of the additional scalar exhibits a behavior relevant to aforementioned issues. In particular we shall discuss selfgravitating monopole solutions in which the $\chi$ field could develop a halo surrounding the monopole core, a problem relevant in connection with dark matter.

The paper is organized as follows. In section 2 we introduce the $SU(2)$ Yang-Mills-Higgs model in curved space, coupled via a Higgs portal to an additional uncharged scalar. Then in section 3 we present monopole solutions to this model in the case  of a Schwarzschild black-hole background. The solutions exhibit a halo produced by the competing $\chi$ scalar. The self-gravitating case is discussed in section 4 showing the different types of solutions depending on the parameter values. Finally in section 5 we summarize and discuss our results.

\section{The model}

The action for the $SU(2)$ Yang-Mills-Higgs model  coupled to gravity in a four dimensional space-time with signature $(-,+,+,+)$, together with a singlet scalar field $\chi$ coupled to the Higgs field reads
\be\label{action}
S = S_G + S_{YM} + S_H + S_\chi =
\int d^4x\,\sqrt{|g|}\; ( {\mathcal L}_G + {\mathcal L}_{YM} + {\mathcal L}_H + {\mathcal L}_\chi)
\ee
where
\ba\label{lagrange}
{\mathcal L}_G &=& \frac{1}{2\,\kappa^2}\,R\cr
{\mathcal L}_{YM}&=& -\frac{1}{4\,e^2}\; F_{\mu\nu}^a F^{a\,\mu\nu}\cr
{\mathcal L}_H&=& -\frac{1}{2}    D_\mu H^a \; D^\mu H^a - V(H)\cr
{\mathcal L}_\chi&=& -\frac{1}{2}\nabla_\mu \chi \;\nabla^\mu \chi - U(\chi, H)
\ea
We use indices $a, b =1,2,3$ for the $SU(2)$ algebra and $\mu  = (0,1,2,3)$ for space-time.
Parameter $\kappa$ is defined as 
$\kappa^2\equiv 8\,\pi\, G_N$
with $G_N$ is the Newton constant,  and $e$  is the gauge coupling.
Concerning the scalar potentials they are chosen as
\be
V(H) = \frac{\lambda}{4}\; ( H^a H^a - h_0{}^2 )^2
\label{VH}
\ee
and
\be
U(\chi, H) = \frac{h_0{}^2\,\alpha}{2}\; \chi^2 + \frac{\beta}{4}\; \chi^4 + \frac{\gamma}{2}\;
H^a H^a
 \; \chi^2
\label{Uchi}
\ee
with $\alpha<0$ and $\lambda$, $\beta$, $\gamma >0$, dimensionless coupling constants.
The field strength   $F^a_{\mu \nu}$ ($a=1,2,3$)  is defined as:
\be
F^a_{\mu \nu} = \nabla_\mu A_\nu^a - \nabla_\nu A_\mu^a +
\varepsilon^{abc}A_\mu^b A_\nu^c \label{defF}
\ee
and  the covariant derivative $D_\mu$ acting on the Higgs
triplet $H^a$ is given by
\be D_\mu H^a = \nabla_\mu H^a + \varepsilon^{abc} A_\mu^b H^c
\label{dercov} \ee
where $\nabla$ stands for usual covariant derivative associated to the metric $g_{\mu\nu}$.

After symmetry breaking from  $SU(2)$  to $U(1)$ gauge group the model describes a  massless gauge field  together with a  massive vector field $W$ with mass $m_W$, the Higgs scalar with mass $m_H$ and the uncharged scalar field with mass
$m_\chi$, with
\be\label{masas}
m_W = e\,h_0\qquad,\qquad  m_H = \sqrt{2\,\lambda}\;h_0\qquad,\qquad
m_\chi= \sqrt{\gamma - |\alpha|}\;h_0
\ee
Note that there won't be any problems with the sign under the square root in $m_\chi$ since we will be taking $\gamma > |\alpha|$, as we will see in Section \ref{sec:vacuum}.

Concerning gravity, the field equations that follow from (\ref{action}) are:
\be\label{eomg}
R_{\mu\nu} - \frac{R}{2}\;g_{\mu\nu} =
\kappa^2 \; \left(T_{\mu\nu}^{YM} + T_{\mu\nu}^H +
T_{\mu\nu}^\chi\right)
\ee
where the energy-momentum tensor
$\;T_{\mu\nu}\equiv -2\,\frac{\delta S}{\delta g^{\mu\nu}}$ has the contributions,
\ba
T_{\mu\nu}^{YM} &=& \frac{1}{e{}^2} F_{\mu \rho}^a
F^a_\nu{}^\rho + g_{\mu\nu}\; L_{YM}\cr
T_{\mu\nu}^H &=&\; D_\mu H^a\; D_\nu H^a + g_{\mu\nu}\;L_H\cr
T_{\mu\nu}^\chi &=&\; \nabla_\mu \chi\; \nabla_\nu \chi + g_{\mu\nu}\;
L_\chi
\label{T}
\ea
In the case of matter, gauge and scalar fields equations are,
\bea\label{eomA}
\frac{1}{e{}^2} D^\rho F^{a}_{\mu \rho} &=&
\varepsilon^{abc}\left( D_\mu H^b\right) H^c
\\
\label{eomH}
D^\rho D_\rho H^a &=& 
\left(\lambda\; ( H^b H^b - h_0{}^2 )
+ \gamma\;\chi^2\right)\;H^a\\
\label{eomchi}
\nabla^\rho \nabla_\rho \chi &=& \frac{\delta U(\chi, H)}{\delta \chi}
= \left(h_0{}^2\,\alpha + \beta\;\chi^2 +\gamma\;(H^b H^b )\;\right)\;\chi
\ea

The most general static, spherically symmetric form for the metric in $3$ spatial  dimensions together with the t'Hooft-Polyakov ansatz for the gauge and scalar fields in the usual vector notation reads,
\ba
g &=& - f(x)\; A(x)^2\; d^2 t + f(x)^{-1}\; d^2 r +r^2\; d^2\Omega_2\nonumber\\
\vec A &=&      - d\theta\; (1 - K(x) )\; \vfi
+ d\varphi \; (1 - K(x) )\; \sin\theta\; \vteta\nonumber\\
\vec H &=& h_0\; H(x)\; \vr\cr
\chi &=& h_0\; \chi(x)
\label{ansatz}
\ea
where we  have introduced the dimensionless radial coordinate $\; x\equiv e\; h_0\; r$
and $(\vr, \vteta,\vfi)$ are the standard spherical unit vectors (see Appendix).
Ansatz (\ref{ansatz}) together with the boundary conditions to be discussed  below  should lead to  solutions magnetically charged under the $U(1)$ 't Hooft-Polyakov field strenght
${\cal F}_{ij}\equiv \frac{1}{e\,h_0}\vec H \cdot\vec F_{ij},\; i,j=1,2,3$, with magnetic charge
\be
Q_m \equiv\int_{{S^2}{\left|_{r\rightarrow\infty}\right.}} {\cal F}= -\frac{4\pi}{e}
\ee
The field equations (\ref{eomg})-(\ref{eomchi}) read,
%
\ba
\left( f (x)\; A(x)\; K'(x) \right)' &=& A(x)\; K(x) \left(
\frac{K(x)^2-1}{x^2} + H(x)^2\right)\cr
\left( x^2\;f(x)\; A(x)\; H'(x) \right)' &=& A(x)\; H(x) \left( 2\; K(x)^2
+\frac{x^2}{e^2}\,\left( \lambda\;( H(x)^2-1) + \gamma\;\chi(x)^2\right)
\right)\cr
\left( x^2\;f(x)\; A(x)\;\chi'(x) \right)' &=&  \frac{x^2}{e^2}\,A(x)\;\chi(x)\;
\left( \alpha + \beta\;\chi(x)^2+\gamma\;H(x)^2\right)\cr
%
 \left( x\; f(x)\right)' &=& 1  -\bar\kappa^2\, \left( f(x)\; V_1 + V_2  \right)\cr
 x\; A'(x) &=&\bar\kappa^2\; V_1 \;A(x)
\label{eqgravmatter}
\ea
where
\ba
V_1 &=& K'(x)^2 + \frac{x^2}{2}\,\left(H'(x)^2 + \chi'(x)^2\right)\cr
V_2 &=& \frac{ (K(x)^2 -1)^2}{2\; x^2} + K(x)^2\;H(x)^2\cr
&+& \frac{x^2}{e^2}\;\left(\frac{\lambda}{4}\;(H(x)^2 -1)^2
+ \frac{\alpha}{2}\; \chi(x)^2 + \frac{\beta}{4}\;\chi(x)^4
+\frac{\gamma}{2}\; H(x)^2\; \chi(x)^2\right)\cr
\nonumber
\ea
Here we have introduced the dimensionless gravitational coupling
$\bar\kappa\equiv h_0\,\kappa$
that together with $(e,\lambda, \alpha, \beta, \gamma)$ defines the parameter space of the theory.
Note that  $\;\bar\kappa =  {(8\,\pi G_N)}^{1/2} h_0\;$ so that, at $G_N$ fixed, its change corresponds to a change of the Higgs vacuum expectation value.

\section{The monopole solution in a black hole background}
In the    absence of back-reaction ($\bar \kappa = 0$)  we shall take a Schwarzschild black hole as a background,
\be
f^{Schw}(x)=1-\frac{x_h}{x}\qquad,\qquad
A^{Schw}(x)=1
\ee
leading to the following matter field equations
\begin{eqnarray}
\left( f^{Schw} (x)\, K'(x) \right)' &=&  K(x)
\left( \frac{K(x)^2-1}{x^2} + H(x)^2
\right) \cr
\left( x^2\;f^{Schw}(x)\; H'(x) \right)' &=& H(x) \left( 2\; K(x)^2
+\frac{x^2}{e^2}\;\left( \lambda\;( H(x)^2-1) + \gamma\;\chi(x)^2\right)
\right)\cr
\left( x^2\;f^{Schw}(x)\;\chi'(x) \right)'&=&
\frac{x^2}{e^2}\;\chi(x)\; \left( \alpha + \beta\;\chi(x)^2+\gamma\;H(x)^2\right)
\label{eommattersinbr}
\end{eqnarray}
Asymptotically we shall impose that the solution goes to the vacuum,
\be
K(\infty) = 0\qquad,\qquad H(\infty)= 1
\qquad,\qquad \chi(\infty)= 0
\label{20}
\ee
It is important to note that in general a given solution of Eqs. \eqref{eommattersinbr} would depend, at fixed coupling constants, on the position of the horizon $x_h$.
However,  $\;f(x_h)=0\;$ makes the system singular, imposing on the matter equations  three constraints on $x_h$.
This leaves us  with three free parameters that are fixed by Eq.\eqref{20}.

Concerning the energy associated to static solutions in the background, it takes the form,
\bea
E &\equiv& \int dr\,r^2\,d\Omega_2\, A(x)\,T_{00}\cr
&=& \!\frac{4 \pi h_0}{e}\!\!\int_{x_h}^\infty\!\! dx\,
A(x)\,\left(\vphantom{\int_{x_h}^\infty } f(x)\,\left(K'(x)^2 + \frac{x^2}{2}\,\left(H'(x)^2 + \chi'(x)^2\right)\right) \right.
+\frac{ (K(x)^2 -1)^2}{2\; x^2}\cr
&+&\left. K(x)^2\;H(x)^2 \right.
+ \left. \frac{x^2}{e^2}\;\left(\frac{\lambda}{4}\;(H(x)^2 -1)^2
+ \frac{\alpha}{2}\; \chi(x)^2 + \frac{\beta}{4}\;\chi(x)^4
+\frac{\gamma}{2}\; H(x)^2\; \chi(x)^2\right)
\vphantom{\int_{x_h}^\infty } \right)\cr
&\equiv& \!\frac{4 \pi h_0}{e}\!\!\int_{x_h}^\infty\!\! dx\, x^2\,A(x)\, \mathcal{E}(x)
\label{energy}
\eea
where we have introduced $\;\mathcal{E}(x)\equiv T_{00}(x) / (e^2\,h_0{}^4)$
as a dimensionless energy density of the system.
It is  easy to see that the solutions of the field equations \eqref{eommattersinbr}  are extrema of this functional.

\bigskip

\subsection{Vacuum Structure} \label{sec:vacuum}

In order to find translationally invariant  vacuum states all derivatives of the profile functions should vanish. The associated values of  $K$,$H$ and $\chi$ can be obtained from  equations  \eqref{eommattersinbr}.
Using the notation $(K^2,H^2,\chi^2)$ we get,
\begin{equation}
{\rm Vacuum~states}\quad
\left\{\begin{array}{rcl}
I&=&(0,0,0)\\
II&=&\left(0,0,-\frac{\alpha}{\beta}\right)\\
III&=&(0,1,0)\\
IV&=&\left(0, \frac{\gamma \alpha + \lambda \beta}{\lambda \beta - \gamma^2},\frac{- \lambda (\alpha + \gamma)}{\lambda \beta - \gamma^2}\right)\\
V&=&(1,0,0)\\
VI&=&\left(1,0,-\frac{\alpha}{\beta}\right)
\end{array}\right.
\label{romanos}
\end{equation}
Since we look for $K$ and $\chi$ solutions  going to zero and  $H$ going  to $1$ asymptotically, the right vacuum results $III$. We then compare vacuum  energy  densities with respect to it in \eqref{romanos}. Having into account that
in the limit of infinite volume only the last term in (\ref{energy}) contributes to $\;\mathcal{E}(x)\;$ we get,
\begin{equation}
{\rm Vacuum~energies}\quad
\left\{\begin{array}{rcl}
\mathcal{E}^I&=& \mathcal{E}^V
= \frac{\lambda}{4}\\
\mathcal{E}^{II}&=& \mathcal{E}^{VI}=\frac{1}{4} (\lambda - \frac{\alpha^2}{\beta})\\
\mathcal{E}^{III}&=&0 \label{T}\\
\mathcal{E}^{IV}&=&\frac{\lambda}{4} \frac{(\alpha+\gamma)^2}{\gamma^2-\lambda \beta}
\end{array}\right.
\end{equation}
Given that $\lambda$ is positive, we find two necessary conditions so that vacuum $III$ is the actual vacuum of the theory. From the requirement that $\mathcal{E}^{II}$ and $\mathcal{E}^{IV}$ are positive, we find
\begin{equation} \label{bounds}
\frac{\alpha^2}{\beta} < \lambda < \frac{\gamma^2}{\beta}
\end{equation}
So, only if $\lambda$ falls in this region we can expect to find solutions of stable  vacuum state $III$ or asymptotic to it.
We  will look for solutions where also the field $\chi$, that goes  asymptotically to zero, takes a constant non trivial value at the horizon, a configuration that can be interpreted as providing a halo to the monopole.

\subsection{Numerical results}
We solved system \eqref{eommattersinbr}  using the relaxation method \cite{art} which determines the solution from an initial guess and improving it iteratively. The natural initial guess is the Prasad-Sommerfield monopole solution \cite{PS} in flat space.

\begin{figure}[h]
\centering
\includegraphics[scale=.2]{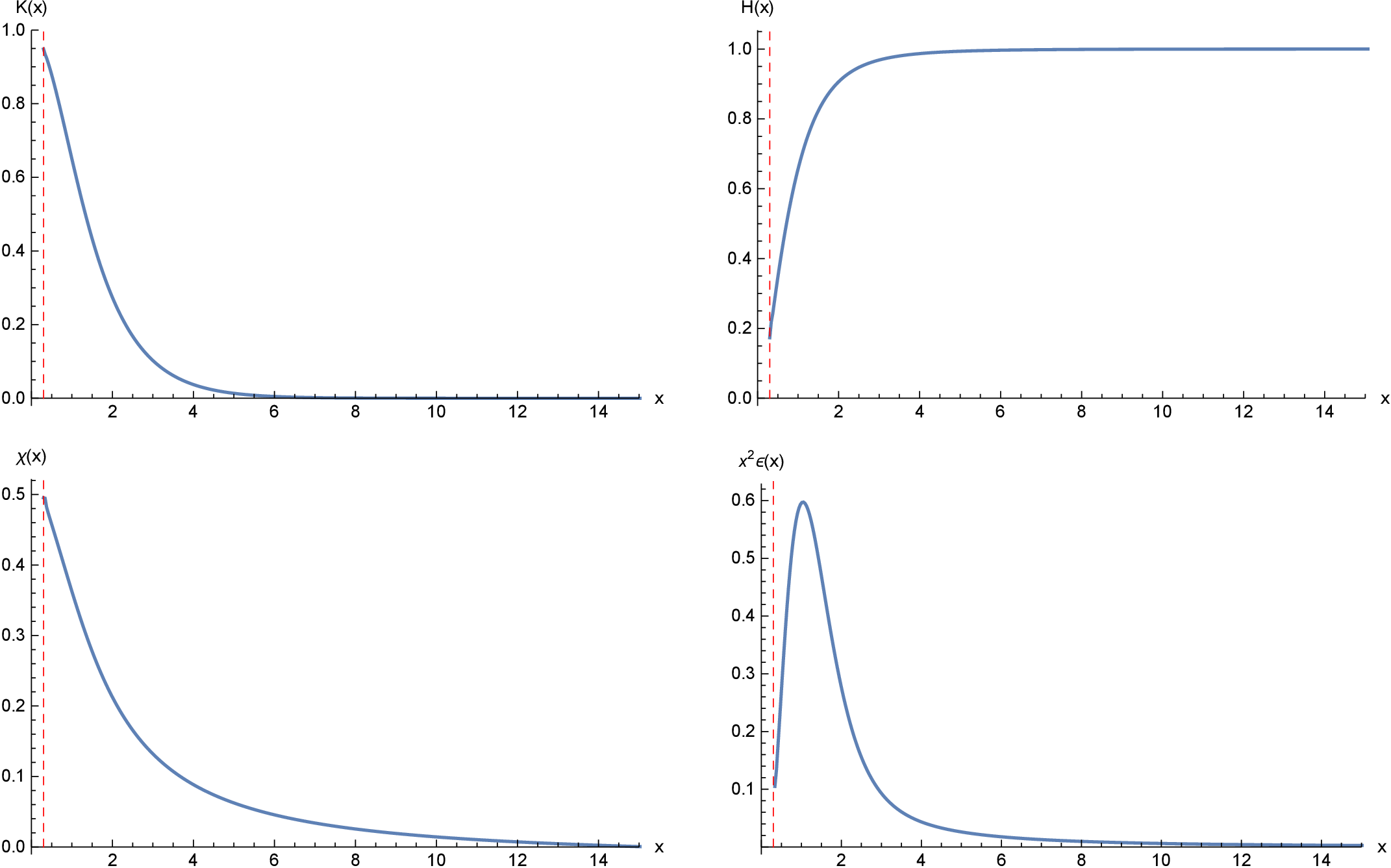}
\caption{Profile functions for $K(x)$, $H(x)$, $\chi(x)$ and the energy density $x^2 \mathcal{E}(x)$. The parameters take the values $x_h=0.3$, $e=0.303$, $\lambda=0.129$, $\alpha = -0.374$, $\beta=1.086$, $\gamma=0.375$. }
\label{fig:ejemploresultados}
\end{figure}

We present in Fig. \ref{fig:ejemploresultados}  profiles of the scalars and magnetic fields and the energy density $\mathcal{E}(x)$, as defined in \eqref{energy}.
Concerning the choice of parameters, in order to have a large halo a small
$\chi$ mass is required. From Eq. \eqref{masas} this implies that $\gamma$ and $|\alpha|$ must be of the same order. Moreover, when exploring different values of $\gamma$ and
$|\alpha|$ looking for large halos we have to keep $\alpha$ and $\gamma$ in view of  inequalities  \eqref{bounds}. As a result of this, although   a change in the maximum value of $\chi(x)$ takes place, the   profiles are similar. The remaining parameters were chosen close to the values in the electroweak scale.

\begin{figure}[H]
\centering{\includegraphics[scale=.2]{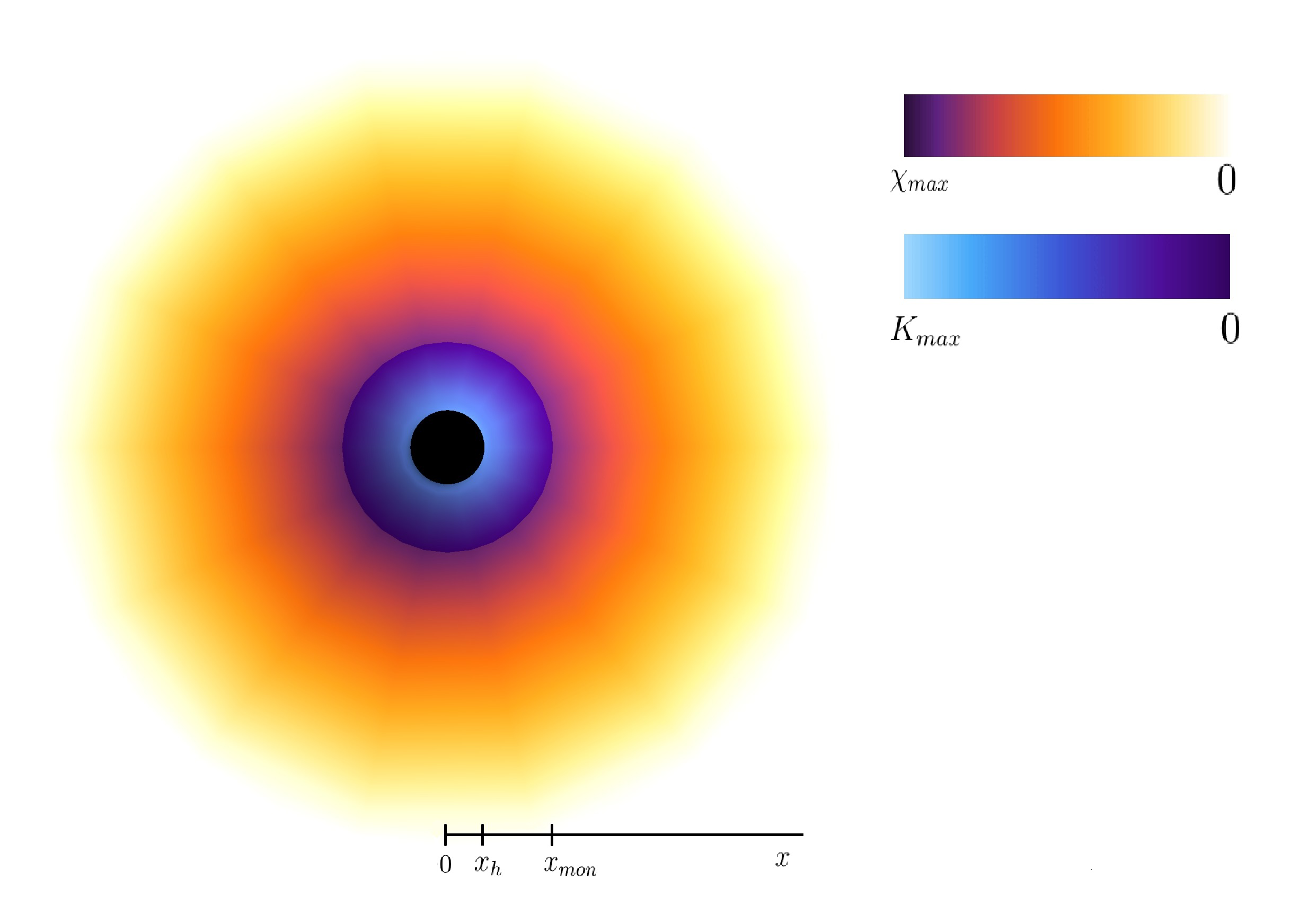}}
\caption{\,\,
{A cross section of the black hole inside the non Abelian monopole (represented here by the gauge field $K$) which in turn is surrounded by the $\chi$ halo, in the  color gradient display.
The figure corresponds to a solution in which  parameters take the values $x_h=0.3$ ($x_{mon} \sim 1$ since $r_{mon} \sim 1/(e h_0)$ \cite{Weinberg}), $e=0.303$, $\lambda=0.129$, $\alpha = -0.374$, $\beta=1.086$, $\gamma=0.375$, like in Fig. \ref{fig:ejemploresultados}, where one can see the corresponding values of $\chi_{max}$ and $K_{max}$.}}
\label{fig:ejemplohalo}
\end{figure}

As it also  happens in the case with no additional competing field $\chi$, the solution we found corresponds to a black hole with its horizon inside a monopole  provided $h_0$ has an upper bound $h_0^{crit}$ 
 since otherwise no stable solutions exists (see \cite{Volkov} and references therein). In our notation since $x = e h_0 r$ this implies a critical value for $x_h$. We have chosen to represent in Fig.\,1 the solutions for  $x_h=0.3$, a value that is smaller than the critical value which is comparable to the monopole radius.
 Under such condition the presence of the $\chi$ field produces a halo around the monopole, as can be seen in
Fig. \ref{fig:ejemplohalo} which shows the cross section of the black hole, the monopole and the $\chi$ field.

\section{The self-gravitating case}

Here we shall look for a solution to equations  \eqref{eqgravmatter} with a horizon at nonzero radius $x = x_h$. The boundary conditions we impose on the metric functions are,
\be \label{CCgrav}
f(x_h) = 0  \qquad;\qquad \lim_{x \to \infty} A(x) = 1
\ee
Concerning gauge and matter fields we have imposed the same conditions as in the no back-reaction case, namely Eq.\,\eqref{20}.

The ``trivial" solution to system \eqref{eqgravmatter} corresponds to the vacuum of gauge and matter fields, that is,
\be
K(x)=0 \qquad , \qquad H(x)=1 \qquad , \qquad \chi(x)=0
\ee
that, together with boundary conditions \eqref{CCgrav}, imply the following behaviour for the metric functions
\be
f(x) = 1- \frac{x_h \left(1+\frac{\bar\kappa^2}{2x_h^2}\right)}{x} + \frac{\bar\kappa^2}{2x^2}\qquad , \qquad A(x)=1
\ee
which corresponds to the well known Reissner-Nordstr\"om (RN) solution. Note that by imposing the condition $f(x_h)=0$ we are discarding the case of naked singularities.

\begin{figure}[h]
\centering{\includegraphics[scale=.15]{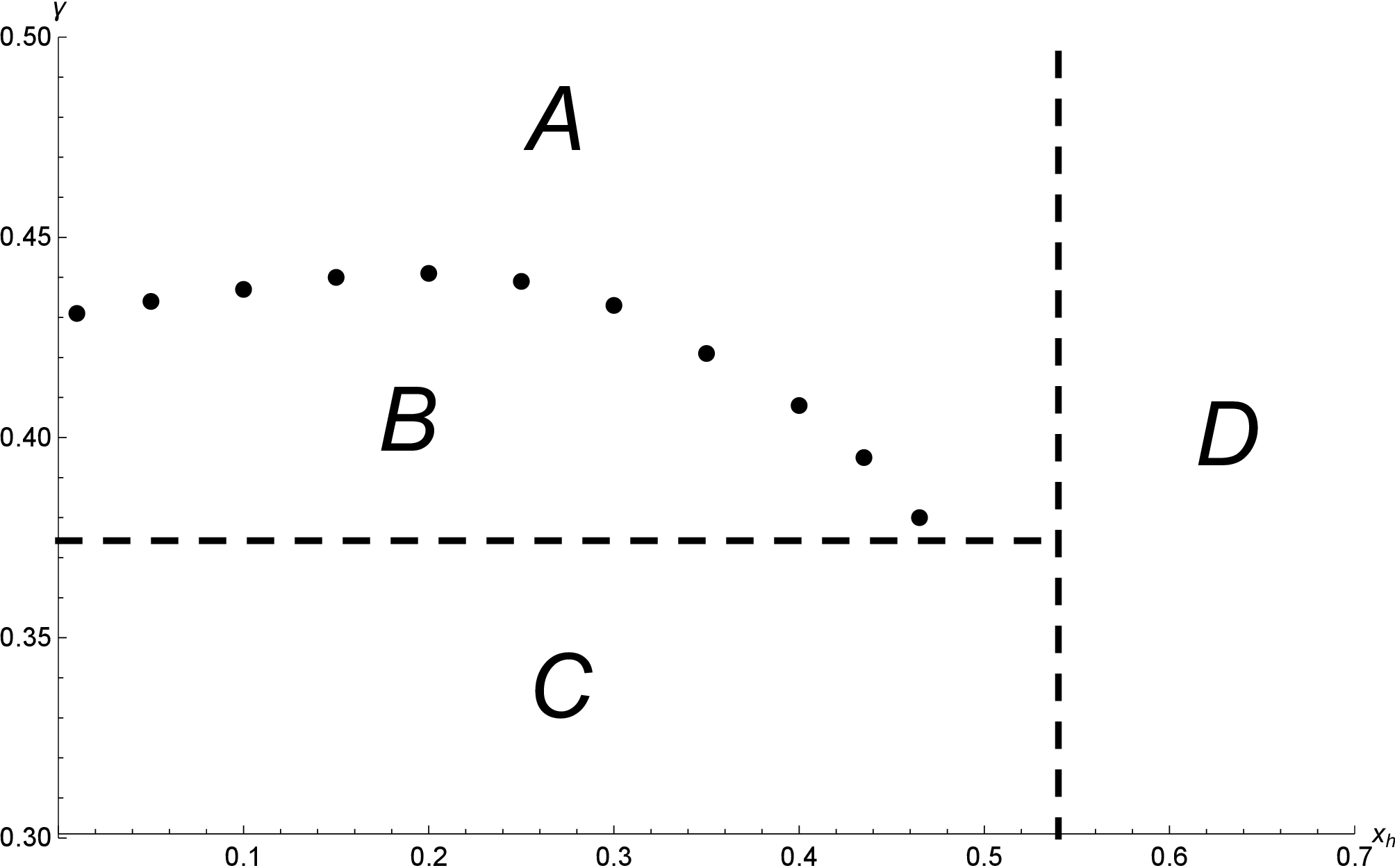}}
\caption{Phase diagram of solutions with fixed parameters for $e=0.303$, $\lambda=0.129$, $\alpha = -0.374$, $\beta=1.086$. Black hole-monopole-halo solutions exist only in region B, and without halo in Region A. Region C is forbidden by the stability bound \eqref{bounds} and in Region D only black hole solutions exist. }
\label{fig:phasediag}
\end{figure}

The numerical results for small $\bar\kappa$ lead to field profiles that are very similar to those in the previous section (Figs. 1 and 2). Moreover, in order to study the possible existence  of solutions,
we have also constructed a zero temperature phase diagram (Fig.\,\ref{fig:phasediag})
in terms of $x_h$ and   $\gamma$. We have taken parameter values $e,\lambda, \alpha, \beta$  as in the previous figures and for $\bar\kappa$ a very small value close to the actual one.

We have found four regions that exhibit a completely different behavior. Region B is the one in which   black-hole-monopole-halo solutions exist and the profiles are like those in Figs.\,\ref{fig:ejemploresultados}-2. For larger values of $\gamma$, in region A,   there is no halo, i.e $\chi = 0$ and we find the  black-hole-monopole solution constructed in \cite{GV}-\cite{Bizon}. Region C is where the bound \eqref{T}  is not satisfied and vacuum $III$ is unstable. Finally when the horizon radius grows and becomes larger    than the monopole radius, the latter is swallowed up and there exists solely the black hole solution (region D). Notice that in Fig.\,\ref{fig:phasediag}  we have chosen certain values for the mixing $\phi\chi$ potential parameters $\alpha$ and $\beta$; other choices lead qualitatively similar diagrams.

~

~

 \begin{figure}[h]
\centering{\includegraphics[scale=.15]{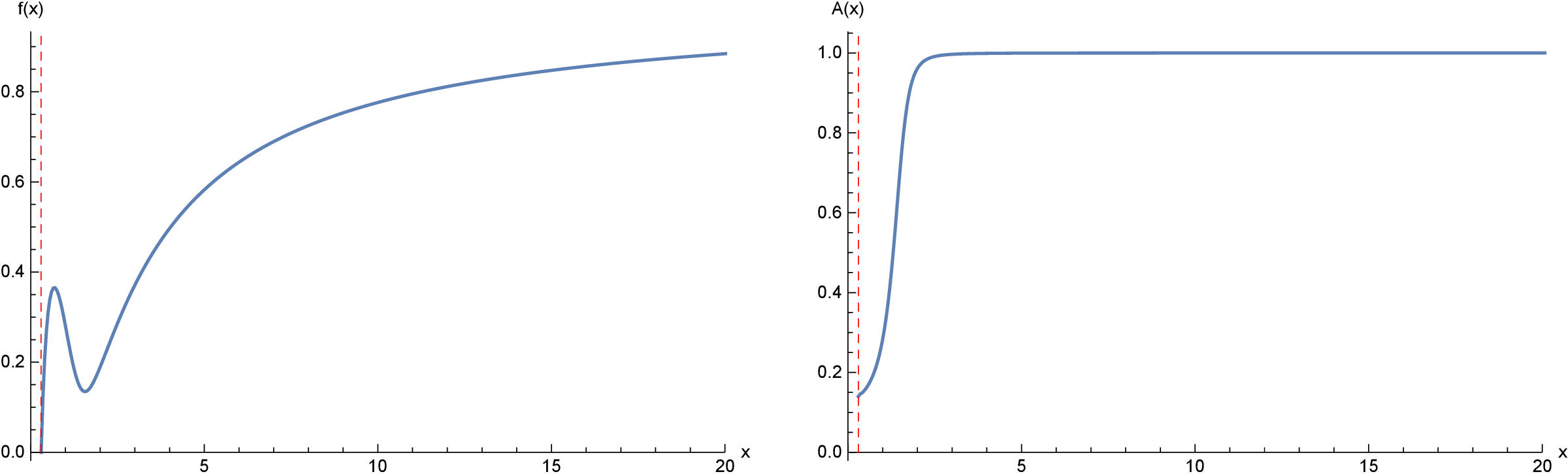}}
\caption{An example of the behavior at large $\bar\kappa$ and the appearance of the second horizon, for parameters $x_h=0.3$, $e=25$, $\bar\kappa=1.6$, $\lambda=6.1$, $\alpha = -9.981$, $\beta=16.333$, $\gamma=9.982$ }
\label{fig:2dohorizonte}
\end{figure}

~

Concerning large $\bar\kappa$ behavior the profiles of the metric function start bending as in  the quasi Schwarzschild black hole cases and finally the minimum of $f(x)$ approaches the axis where a second horizon appears (as was previously discussed in \cite{LMS1}).   This can be seen from an analysis of the metric profile functions before this critical point. This behavior is depicted in  Fig. \ref{fig:2dohorizonte}.

\section{Summary and discussion}
In this work we have studied non-Abelian 't Hooft-Polyakov solutions in curved space in the presence of an additional uncharged scalar coupled to the visible  Georgi-Glashow model through a Higgs portal.
We have first solved numerically the equations for the  gauge-matter model  in a Schwarzschild black hole background using the  relaxation method. In this way we established coupling constant bounds in order to have
 stable black hole-monopole   solutions surrounded by a $\chi$ halo as the one shown in Fig.\,2.
In order to have a large halo a small
$\chi$ mass is required. This implies from \eqref{masas} that $\gamma$, the  Higgs-$\chi$ coupling constant and $\alpha$, the parameter that appears in the  quadratic $\chi^2$ term should be of the same order.

We then  considered the self-gravitating case for which we were able to establish a phase diagram where one can identify four regions, depending on the mixing coupling constant $\gamma$ and the horizon radius $x_h$. For large values of $x_h$ the black hole absorbs the monopole and no non-trivial black hole-monopole solutions exist. To the other side of this critical line three regions of distinct solutions can be identified according to the value of $\gamma$. For very small values of the mixing parameter, the bound \eqref{bounds} is not satisfied and the solution with a halo is unstable. For very large values of $\gamma$ we find the well-known black hole-monopole solutions with $\chi=0$. Only for values of $\gamma$ between these regions do we encounter the non-trivial black hole-monopole-halo solutions. This is, we think, an important lesson from our results:  for large values of the mixing parameter $\gamma$ no $\chi$ halo exists and the solution reduces to the black hole-monopole solution \cite{GV}-\cite{Volkov} and, moreover, the  $\chi$ field  presence imposes a bound for the Higgs potential coupling constant through the inequality \eqref{bounds}, properties that agree with experimental constraints in models in which a mixing between visible and hidden sectors are discussed in connection with the dark matter problem.

In view of the discussion above, we would like to make a final comment  on a possible connection of our results with models in which dark matter is composed of (pseudo) scalars that could coalesce into halos. A first condition for the solution that we present is satisfied, namely that the existence of black holes inside monopoles is consistent with scales of  grand unified theories \cite{Preskill}. Concerning the $\chi$ scalar mass, according to eq.\,\eqref{masas} it depends on the difference between $\gamma$   and $\alpha$ parameters introduced in the $U(\chi,H)$  potential (4). In particular the $\gamma -|\alpha|$ difference can be chosen to be of the order of mass scales of the (pseudo) scalars in dark matter models. Indeed, keeping the Higgs portal coupling constant sufficiently small, one can adjust the value of the parameter of the $\chi^2$ term so that the $\chi$ mass can take values like those   of the (pseudo) scalar dark mass, of the order $10^{-22}$\,eV    as in refs.\,\cite{schive}-\cite{Helfer} or those more recently proposed in ref.\,\cite{Smoot}, with a  mass of the order $10^{-18}$\,eV.

\subsection*{\bf Acknowledgements}
F.A.S. would like to thank Paola Arias for helpful comments. The work of F.A.S. is supported by CONICET grant PIP 688  and FONCYT grant PICT 2304. The work by A.L. is supported by CONICET grant PIP 688.

\end{document}